# Quantum dynamics of proton migration in $H_2O$ dications: formation of $H_2^+$ on ultrafast timescales

Manish Garg,[a] Ashwani K. Tiwari,[a,*] and Deepak Mathur[b]

[a] Indian Institute of Science Education and Research Kolkata, Mohanpur 741 252, India

[b] Tata Institute of Fundamental Research, 1 Homi Bhabha Road, Mumbai 400 005, India

**Abstract**

Irradiation of isolated water molecules by few-cycle pulses of intense infrared laser light can give rise to ultrafast rearrangement resulting in formation of the $H_2^+$ ion. Such unimolecular reactions occur on the potential energy surface of the $H_2O^{2+}$ dication that is accessed when peak laser intensities in the $10^{15}$ W cm$^{-2}$ range and pulse durations as short as 9-10 fs are used; ion yields of ~1.5% are measured. We also study such reactions by means of time-dependent wavepacket dynamics on an *ab initio* potential energy surface of the dication and show that a proton, generated from O-H bond rupture, migrates towards the H-atom, and forms vibrationally-excited $H_2^+$ in a well-defined spatial zone.

[*] E-mail: ashwani@iiserkol.ac.in

**I Introduction**

It is now well established that irradiation of gas-phase molecules by femtosecond-long laser pulses of peak intensity in excess a few TW cm$^{-2}$ results in multiple ionization and subsequent fragmentation (for a compilation of cogent reviews, see Ref. 1, and references therein). In such studies, the magnitude of the optical field matches the intra-molecular Coulombic field. As a result, the laser-molecule interaction is strongly dominated by single and multiple ionization. The field-induced removal of one or several electrons from the irradiated molecule subsequently results in the breaking of one or several bonds. In the course of the last seven years or so, experiments have begun to reveal that in addition to the strong optical fields inducing bond breakages in multiply charged molecules, formation of bonds may also be induced.[2-5] Such bond formation usually involves a unimolecular process in which intra-molecular migration of protons or hydrogen atoms occurs on ultrafast timescales. This has opened possibilities of new molecular entities being formed in the strong field regime.[6,7]

Experimental work on bond formation processes induced by strong fields of femtosecond duration has almost exclusively focused attention on hydrocarbon molecules.[2-8] However, very recently we reported preliminary results of strong field experiments on a triatomic molecule, $H_2O$ which, upon exposure to few-cycle pulses (of 9 fs duration) of peak intensity in the $10^{15}$ W cm$^{-2}$ range, underwent fast rearrangement so as to produce the $H_2^+$ ion.[9] Although quantumchemical calculations accompanying this experimental work indicated that there exists a $H_2O^{2+}$ dication state which asymptotically leads to fragments that include $H_2^+$, it was not possible to obtain proper insights into the dynamics. We present here results of a time-dependent quantum dynamics study of the $H_2O^{2+}$ dication that supplement our experimental findings. Specifically, we generate, *ab initio*, a potential energy surface (PES) of the ground electronic state of $H_2O^{2+}$;

we solve the three-dimensional time dependent Schrödinger equation in Jacobi coordinates for an initial wavepacket (which is the ground vibrational state wave function of neutral $H_2O$) and follow its time evolution for a large number of small time steps on the $H_2O^{2+}$ PES. The results of our study indicate that a proton, generated upon the breaking of the O-H bond, migrates towards the H-atom, thus forming $H_2^+$ in a well-defined spatial zone within the dication PES. We calculate the norm of the wavepacket entering this zone and obtain a finite probability in excess of 1.4% for the $H_2^+$ channel. This result is found to be in consonance with the experimental findings of $H_2^+$ ion yield that we report here. We also find that the average H-H separation in the $H_2^+$ ion formed in this process is higher than the equilibrium bond length, indicating that the ion that is formed is vibrationally excited.

**II Methodology**

**A. Experimental studies.** Almost all earlier work on few-cycle ionization dynamics in molecules in the strong field regime has used the hollow-fibre pulse compression technique.[10] Recently, few-cycle pulses have also been generated using filamentation in gas-filled tubes.[11,12] In our experiments, we used 0.4 mJ, 40 fs long pulses of 800 nm light (at 1 kHz repetition rate) from a Ti-sapphire amplifier to generate few-cycle pulses (Fig. 1). After passing through an aperture (A1) the laser beam was focused with a 1.0 m focal length metal coated spherical mirror on to a 1.5 m long tube filled with Ar gas at 1.2 atm pressure. The central part of the resulting broadband light was selected using a slit (A2) and was temporally compressed by means of a set of chirped dielectric mirrors (CDM1) so as to yield pulses of 15 fs duration with an energy of 0.25 mJ. These pulses were then passed through a 1 m long second tube filled with 0.9 atm of Ar gas and the resulting broadband light was again temporally compressed using a second pair of chirped dielectric mirrors (CDM2) so as to yield 10 fs (four-cycle) pulses with typical energy values of 0.25 µJ.

We note here an important advantage of the dual tandem-cell method used in our experiments is that intensity clamping in the filaments results in significant reduction in energy fluctuations in the laser beam.[12] Such stability enhancement becomes important in studies involving molecules.

We characterized the compressed laser pulses by means of spectral phase interferometry for direct electric field reconstruction (SPIDER). These pulses were steered through a 300 µm thick fused silica window into an ultrahigh vacuum (UHV) chamber in which the laser-$H_2O$ interaction occurred at pressures that were low enough (~$10^{-9}$ Torr) to obviate the need to consider either space charge saturation or propagation effects. The few-cycle pulses were precompensated for the chirp and were focused within the UHV chamber by a spherical mirror ($f$ = 5 cm).

Ion analysis was by conventional linear time-of-flight (TOF) methods but we exercised care to ensure that, by judicious use of apertures at the entrance to the TOF spectrometer tube, only those ions that originated in the central, most-intense portion of the laser focal volume were efficiently extracted into our spectrometer. We confirmed that use of ion extraction fields of ~100-200 V cm$^{-1}$ allowed us to achieve nearly unit ion extraction efficiency for ions possessing less than ~5 eV kinetic energy.

**B. Potential Energy Surface.** The ground singlet-state potential energy surface for $H_2O^{2+}$ was generated by using the multireference configuration interaction (MRCI) method with the correlation-consistent polarized-valence quadruple-ζ (cc-pVQZ) basis set. The MOLPRO suite of programs[13] was used on a grid of points with O-H bond lengths $r_{12}$ and $r_{13}$ ranging from $1.6a_0$ to $6.5a_0$ and H-O-H bond angle of $\theta$ = 10º (20º, 30º, 45º, 60º, 90º, 110º, 120º) 150º. An analytical functional fit to the computed *ab initio* potential energy values was obtained using a many-body expansion method.[14,15] The potential energy function for a triatomic system *ABC* (where, in our case, *A*=H, *B*=O, *C*=H) is written as

$$V_{ABC}(R_{AB},R_{BC},R_{AC}) = V^{(1)}_A + V^{(1)}_B + V^{(1)}_C + V^{(2)}_{AB}(R_{AB}) + V^{(2)}_{BC}(R_{BC}) + V^{(2)}_{AC}(R_{AC}) + V^{(3)}_{ABC}(R_{AB},R_{BC},R_{AC}) \quad (1)$$

The diatomic potential for *AB* is denoted by

$$V_{AB} = \frac{C_0 \exp(-\alpha_{AB} R_{AB})}{R_{AB}} + \sum_{i=1}^{L} c_i \rho_{AB}^i \quad (2)$$

where $L=9$ for $OH^+$ and $H_2^+$. The Rydberg type variables $\rho$ are given by

$$\rho_{AB} = R_{AB} \exp(-\beta^l_{AB} R_{AB}), \; l = 2 \text{ (two-body) or 3 (three-body)} \quad (3)$$

Similar expressions hold for *BC* and *CA*.

The linear parameters $c_i$, $i = 0 \ldots N$ and the nonlinear parameters $\alpha_{AB}$ and $\beta_{AB}$ were determined by fitting to the *ab initio* values corresponding to the diatomic fragments (plus a separated atom/ion at a large enough distance) computed with the same basis set used for the triatomic system and using the same *ab initio* procedure. The triatomic term is given by

$$V^{(3)}_{ABC}(R_{AB},R_{BC},R_{AC}) = \sum_{ijk}^{M} d_{ijk} \rho_{AB}^i \rho_{BC}^j \rho_{AC}^k \quad (4)$$

The indices *i*, *j* and *k* vary from zero to a maximum value (*M*) such that

$$i + j + k \neq i \neq j \neq k \tag{5}$$

$$i + j + k \leq M \tag{6}$$

A total of 1069 calculated *ab initio* points were used for the fit that uses $M = 7$. The energy of the three separated atoms/ions in their ground state has been taken to be the zero of the energy. The fitted surface has a root-mean-square deviation of 268 meV compared to the ab initio potential energy values.

Parameters resulting from the analytic fit are listed in Table1. The FORTRAN subroutine describing the PES is available on request.

**C. Time Dependent Quantum Mechanical Methodology**. The three-dimensional interaction Hamiltonian for the $H_2O^{2+}$ in Jacobi coordinates $(R, r, \gamma)$ and for total angular momentum $J = 0$ in the body-fixed Jacobi co-ordinates is given by[16]

$$\hat{H} = \tfrac{1}{2}\left[\frac{P_R^2}{M} + \frac{P_r^2}{\mu}\right] + \frac{\hat{j}^2}{2I} + V(R,r,\gamma)$$

$$= -\frac{\hbar^2}{2}\left[\frac{1}{M}\frac{\partial^2}{\partial r^2} + \frac{1}{\mu}\frac{\partial^2}{\partial R^2}\right] - \frac{\hbar^2}{2I}\frac{1}{\sin\gamma}\frac{\partial}{\partial\gamma}\left(\sin\gamma\frac{\partial}{\partial\gamma}\right) + V(R,r,\gamma) \tag{7}$$

where $P_R$ and $P_r$ are the momentum operators corresponding to the two Jacobi distances $R$ (distance of the O-atom from the center-of-mass of $H_2$) and $r$ (the $H_2$ inter-nuclear distance), respectively, and $\gamma$ is the angle between them. **j** is the rotational angular momentum operator of $H_2$. The quantity M is the reduced mass of (O, $H_2$) and μ is the reduced mass of $H_2$ [M = $m_1(m_2+m_3)/(m_1+m_2+m_3)$ and $\mu = m_2m_3/(m_2 + m_3)$, where $m_1$ is the mass of O and $m_2$ and $m_3$ are the masses of the H atom].

Under our experimental conditions (the molecule being under the influence of ultrashort, intense laser pulses), the transition from $H_2O$ to $H_2O^{2+}$ is assumed to be a Franck-Condon transition and, therefore, the initial wave function $\Psi$ ($t$=0) on the $H_2O^{2+}$ dication PES is taken to be the ground vibrational wave function of neutral $H_2O$ which was generated in the symmetry coordinates and then transformed to the local modes of $H_2O$.[17] The time evolution of $\Psi$ ($t$=0) on the dication PES was carried out by using eqn. 8:

$$\psi(t) = \exp\left[\frac{-i\hat{H}t}{\hbar}\right]\psi(t=0) \qquad (8)$$

The time evolution of the wave packet was carried out using the split-operator algorithm[18] for a large number ($N_T$) of small time steps ($\Delta t$) using the equation

$$\exp\frac{-iH\Delta t}{\hbar} = \exp\frac{-iV\Delta t}{2\hbar}\exp\frac{-ij^2\Delta t}{4I\hbar}\exp\frac{-iT\Delta t}{\hbar}\exp\frac{-ij^2\Delta t}{4I\hbar}\exp\frac{-iV\Delta t}{2\hbar} + O(\Delta t^3) \qquad (9)$$

where $T = P_R^2/2M + P_r^2/2\mu$ is the total radial kinetic energy operator. The action of the exponential operator, $\exp(-iT\Delta t/\hbar)$ is determined using the fast Fourier transform (FFT) method.[19] A discrete variable representation (DVR) method[20-23] is used to the action of the exponential containing the rotational kinetic energy operator, $\exp(-ij^2\Delta t/4I\hbar)$ on the wave function. In the latter case, the eigenfunctions of the $\mathbf{j}^2$ operator, the Legendre polynomials $P_j(\cos\theta_n)$ are used as the finite basis representation (FBR) in which this operator is diagonal. The grid points in DVR are chosen as the nodes of a 29-point Gauss-Legendre quadrature (GLQ).[24] The DVR method involves transforming the wave function to FBR, multiplying it by the diagonal value of this operator $\exp[-ij(j+1)\Delta t \hbar / 4I]$ and transforming it back to the grid representation. Numerically this is accomplished in a single step as follows:

$$\exp[-\frac{ij^2\Delta t}{4I\hbar}]\psi_{lmn'} = \sum_n \sum_j (T^+_{n',j} \exp[-ij(j+1)\Delta t\hbar/4I]T_{j,n})\psi_{lmn'} \tag{10}$$

where $j$ is the rotational quantum number of $H_2$. The coefficients $T_{j,n}$ are the elements of the DVR-FBR transformation matrix constructed in terms of the Legendre polynomials:

$$T_{j,n} = \sqrt{w(n)}\sqrt{\frac{2j+1}{2}}P_j(\cos\theta_n) \tag{11}$$

where $w(n)$ is the weight of the GLQ associated with the point $n$. $T^+_{n,j}$ are the elements of the inverse transformation.

To avoid reflection from the grid edges by the fast moving components of the wave packet, the wave packet at each time step is multiplied by a damping function[25]

$$f(X_i) = \sin\left[\frac{\pi(X_{mask} + \Delta X_{mask} - X_i)}{2\Delta X_{mask}}\right], \quad X_i \geq X_{mask} \tag{12}$$

which is activated in the asymptotic $R$ and $r$ channels. $X_{mask}$ ($X=R,r$) is the point at which the damping function is initiated and $\Delta X_{mask} = X_{max} - X_{mask}$ is the width of X over which the function decays from 1 to 0. The grid parameters are listed in Table 2.

**III RESULTS AND DISCUSSION**

A typical $H_2^+$ ion peak obtained in our time-of-flight spectrum is depicted in Fig. 1. As already noted,[9] molecular ionization is found to totally dominate the mass spectrum that is obtained in this few-cycle temporal regime in the $10^{15}$ W cm$^{-2}$ intensity range. This is in contrast to the situation that prevails when 40–100 fs pulses are used wherein (i) atomic fragments, $O^{q+}$ ($q$=2–5), are observed at yield levels as high as 5%–10% with respect to the yield of the parent molecular ion, $H_2O^+$ at similar intensity values and (ii) molecular fragments, $OH^+$, are produced at yield levels of *2%–5%*. Careful measurements in the $10^{15}$ W cm$^{-2}$ intensity range has enabled us to determine that the integrated ion counts under $H_2^+$ peak in our spectrum (Fig. 1) amount to ~2% of the yield of parent $H_2O^+$ ions. The challenge for our theoretical study is to rationalize, both qualitatively and quantitatively, the observation of $H_2^+$ ions at such yield levels. The initial electronic structure calculations reported in our preliminary communication[9] suggested that $H_2^+$ formation takes place on the $H_2O^{2+}$ PES. Therefore, we have carried detailed electronic structure as well as dynamical calculations for $H_2O^{2+}$, focusing on the dissociation channel that leads to the formation of $H_2^+$ and $O^+$.

There are two possible pathways for the formation of $H_2^+$: (i) the strong field breaks one of the O-H bonds and the resulting proton migrates towards the H-atom, leading to the formation of $H_2^+$, and (ii) both the O-H bonds break simultaneously and the $H^+$ and H fragments approach each other such that $H_2^+$ occurs. To understand the energetics of these two pathways, we plot in Fig. 2 the energy profiles when only one O-H bond is elongated; Fig. 3 shows the corresponding plot when both O-H bonds are simultaneously elongated. It becomes clear from these plots that, for small values of bond angle, there is no potential barrier when only one O-H bond dissociates; there is a small barrier when the bond angle assumes somewhat larger values. In contrast, simultaneous dissociation of both the O-H bonds results in a potential barrier that is distinctly higher. Therefore, it appears unlikely that formation of $H_2^+$ takes place via path (ii). However, in order to develop deeper insights into how the $H_2^+$ forms, we also study the time evolution of the wave packet on the $H_2O^{2+}$ PES. To this end we plot snapshots of the wave packet at different times in Fig. 4 for Jacobi angles $\gamma = 90^0$ and $175^0$.

It is clear from data shown in Fig. 4 that as time progresses, the wave packet moves in a region of the PES where $r$ is very large and $R$ is nearly equal to its initial equilibrium value. This means that the wave packet has moved towards the $OH^+ + H^+$ asymptotic channel. But after some further time, a part of this wave packet starts to move towards the region where $r$ is small and $R$ is large. This region corresponds to the $H_2^+ + O^+$ asymptotic channel. Such dynamics indicate that once one of the O-H bonds is sufficiently elongated, there is a definite probability of $H^+$ moving towards the H-atom and, concurrently, the $O^+$ ion moving away from the H-atom. Such movement also leads to the formation of $H_2^+$.

Data shown in Fig. 4 also makes clear that the probability of formation of $H_2^+$ is large for larger values of $\gamma$. The rectangular region on the dication PES where $r < 3.0$ $a_0$ and $R > 2.0$ $a_0$ can be considered to demarcate the $H_2^+$ formation zone. The norm of the wave packet in this region is a measure of the probability for the formation of $H_2^+$ from double ionization of $H_2O$ to $H_2O^{2+}$ by an ultrashort laser pulse.

The norm of the wave packet in the $H_2^+$ channel at different times is plotted in Fig. 5. It is clear that the formation of $H_2^+$ starts around 10 fs and stops by 15 fs. Indeed, the probability for $H_2^+$ apparently starts decreasing after 15 fs; this is an artifact that arises because we have put a damping function near the edges of the grid which absorbs the wave packet. We deduce from our results that the probability for $H_2^+$ formation on the dication PES is ~1.4%, in excellent accord with our experimental observations.

To gain better understanding of the average value of the bond length of $H_2^+$ that is formed in this process, we have plotted the average value of r ($<r>=<\psi_r|r|\psi_r>/<\psi_r|\psi_r>$, where $\psi_r$ is the part of wave function going into the $H_2^+$ channel) as a function of time in Fig. 6. It is clear that $<r>$ remains nearly constant around 2.6 $a_0$, slightly larger than equilibrium $H_2^+$ bond length, implying that the ion, on average, lies in a higher vibrational state.

In summary, we have carried out time-dependent quantum mechanical calculations on an *ab-initio* potential energy surface of the $H_2O^{2+}$ dication. Our calculations have revealed that there is a finite probability of $H_2^+$ formation when $H_2O$ is doubly ionized. Our calculations also indicate that $H_2^+$ yield would be expected to be about 1.4% relative to the yield of the parent molecular ion, $H_2O^+$, in excellent accord with our experimental results.

**Acknowledgements**
We gratefully acknowledge the skillful contributions of Aditya K. Dharmadhikari, Jayashree A. Dharmadhikari, and Firoz A. Rajgara to the development of the few-cycle laser system.

TABLE 1: Parameters of the $H_2O^{2+}$ potential energy surface, in Å.

Two-body terms for $OH^+$

| | |
|---|---|
| α | .1647666E+01 |
| β | .1500004E+01 |
| $c_0$ | .145607001804E+01 |
| $c_1$ | -.163287059361E+02 |
| $c_2$ | .904603684461E+03 |
| $c_3$ | -.276930561282E+05 |
| $c_4$ | .489123941445E+06 |
| $c_5$ | -.529916818428E+07 |
| $c_6$ | .356701367155E+08 |
| $c_7$ | -.145359023754E+09 |
| $c_8$ | .328112402470E+09 |
| $c_9$ | -.314774282614E+09 |

Two-body terms for $H_2^{\pm}$

| | |
|---|---|
| α | .2133776E+01 |
| β | .1500048E+01 |
| $c_0$ | .420334343648E+00 |
| $c_1$ | -.334158589120E+01 |
| $c_2$ | .146288311779E+03 |
| $c_3$ | -.750347886783E+04 |
| $c_4$ | .228177276116E+06 |
| $c_5$ | -.416013141165E+07 |
| $c_6$ | .458500564683E+08 |
| $c_7$ | -.298053181379E+09 |
| $c_8$ | .104816874633E+10 |
| $c_9$ | -.153367029429E+10 |

Three-body terms $V^{(3)}_{HOH^{2+}}$ for $M = 7$

$\beta_{OH}$ = .1381763437017118E+01   $\beta_{HH}$ = .6426115892327892E-01

| $i$ | $j$ | $k$ | $d_{ijk}$ |
|---|---|---|---|
| 1 | 1 | 0 | .1105098601435149E+02 |
| 1 | 0 | 1 | .4546779885614446E+03 |
| 1 | 1 | 1 | -.6985716315836826E+03 |
| 2 | 1 | 0 | -.1150059877201875E+03 |
| 2 | 0 | 1 | -.4080585878556618E+04 |
| 0 | 2 | 1 | -.1255871447246243E+01 |
| 2 | 1 | 1 | .4821694727141322E+04 |
| 1 | 2 | 1 | .3137267425098153E+03 |
| 2 | 2 | 0 | .7686903512856414E+01 |
| 2 | 0 | 2 | .1141308418650545E+05 |
| 3 | 1 | 0 | .7327835118856301E+03 |
| 3 | 0 | 1 | .2586351887938284E+05 |
| 0 | 3 | 1 | -.2651968329933581E+01 |
| 2 | 2 | 1 | -.1896244256376629E+04 |
| 2 | 1 | 2 | -.1123929388706818E+05 |
| 3 | 1 | 1 | -.1876027336171468E+05 |
| 1 | 3 | 1 | -.4196009534039241E+02 |
| 3 | 2 | 0 | .4813318222204666E+02 |
| 3 | 0 | 2 | -.4337951144220270E+05 |
| 0 | 3 | 2 | .9143502562171109E+01 |
| 4 | 1 | 0 | -.3436327078415850E+04 |
| 4 | 0 | 1 | -.8564677212015141E+05 |
| 0 | 4 | 1 | .1143291801041161E+01 |
| 2 | 2 | 2 | .4300518774824090E+04 |
| 3 | 2 | 1 | .4405329874136921E+04 |
| 3 | 1 | 2 | .1730950181005677E+05 |
| 1 | 3 | 2 | .2726391753039829E+03 |
| 3 | 3 | 0 | -.4465013139856134E+02 |
| 3 | 0 | 3 | .7250601389307820E+05 |
| 4 | 1 | 1 | .3760511176316535E+05 |
| 1 | 4 | 1 | -.1880374217717758E+01 |
| 4 | 2 | 0 | -.4411191821910126E+02 |
| 4 | 0 | 2 | .1042631847150019E+06 |
| 0 | 4 | 2 | -.1884390842314757E+01 |
| 5 | 1 | 0 | .8716959190817190E+04 |
| 5 | 0 | 1 | .1368093065682296E+06 |
| 0 | 5 | 1 | -.1766158873980050E+00 |
| 3 | 2 | 2 | -.3402077524389884E+04 |
| 2 | 3 | 2 | -.4085927734289861E+03 |
| 3 | 3 | 1 | -.2968320274532859E+03 |
| 3 | 1 | 3 | -.8172267106698259E+03 |
| 4 | 2 | 1 | -.4006503486906202E+04 |
| 4 | 1 | 2 | -.1653058006623675E+05 |
| 1 | 4 | 2 | -.1242490818266029E+02 |
| 4 | 3 | 0 | .3723259425669126E+02 |
| 4 | 0 | 3 | -.7925619962626767E+05 |
| 0 | 4 | 3 | .4343220108048946E+01 |
| 5 | 1 | 1 | -.2955412881586891E+05 |
| 1 | 5 | 1 | .5576201768075820E+00 |
| 5 | 2 | 0 | -.2747155498556242E+02 |
| 5 | 0 | 2 | -.8880834071501456E+05 |
| 0 | 5 | 2 | .7874423444940559E-01 |
| 6 | 1 | 0 | -.9038821149544299E+04 |
| 6 | 0 | 1 | -.8036700348401041E+05 |
| 0 | 6 | 1 | .9607287033866570E-02 |

TABLE 2: Grid parameters used in the present study

| Parameters | Values | |
|---|---|---|
| $(N_R, N_r, N_\theta)$ | (64, 64, 29) | Number of grid points |
| $(R_{min}, R_{max})/a_0$ | (0.2, 6.6) | Extension of grid along $R$ |
| $(r_{min}, r_{max})/a_0$ | (1.0, 7.4) | Extension of grid along $r$ |
| $(R_{mask}, r_{mask})/a_0$ | (5.8, 5.8) | Starting point of masking function$(r,R)$ |
| $\Delta t$/as | 24.4 | Time step used in propagation |

**Figure Captions**

Fig. 1. Schematic representation of the apparatus used to generate few-cycle pulses in the present experiments (see text). 10 fs long pulses of 150 µJ energy were focused using a 5 cm curved mirror located within an ultrahigh vacuum chamber so as to yield peak laser intensities in the $10^{15}$ W cm$^{-2}$ range at focused spot where the laser-molecular interaction occurred. Ions formed in this interaction were accelerated into a 20 cm long time-of-flight (TOF) spectrometer. A typical TOF spectrum showing $H_2^+$ ions from $H_2O$ is shown. Also shown are typical traces of the laser pulse duration and the phase, as deduced by SPIDER (see text).

Fig. 2. A section through the fitted potential energy surface of the $H_2O^{2+}$ dication along with *ab-initio* data points at $\theta$ = 30º, 60º, 90º and 120º for $r_{12}$ = 2.0 au. The inset illustrates the coordinate system used for our PES calculation.

Fig. 3. A section through the $H_2O^{2+}$ dication potential energy surface along with *ab-initio* data points at $\theta$ =30º, 60º, 90º and 120º with $r_{12}$ (au) and $r_{13}$ (au) changing simultaneously.

Fig. 4. Three-dimensional probability density plots for the wave packet at different time intervals for $\gamma$ = 90º ( left) and $\gamma$ = 175º (right) during evolution for an initial wave packet at $t$ = 0. Inset shows the Jacobi coordinate system.

Fig. 5. Norm calculated for the evolving wave packet in the $H_2^+$ channel ($r$ < 3.0 au and $R$ > 2.0 au).

Fig. 6. Expectation value of $r$ and $R$ calculated at each time step in the $H_2^+$ channel.

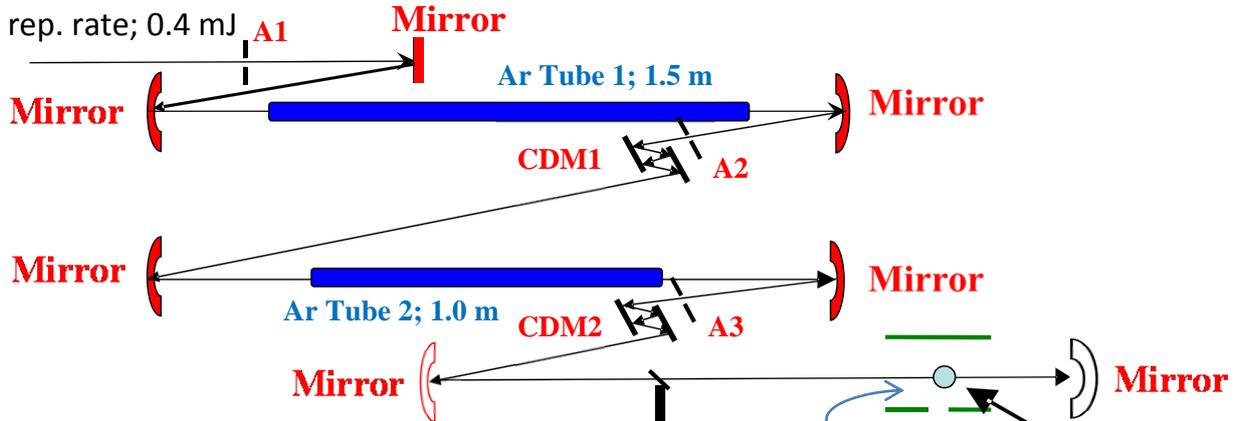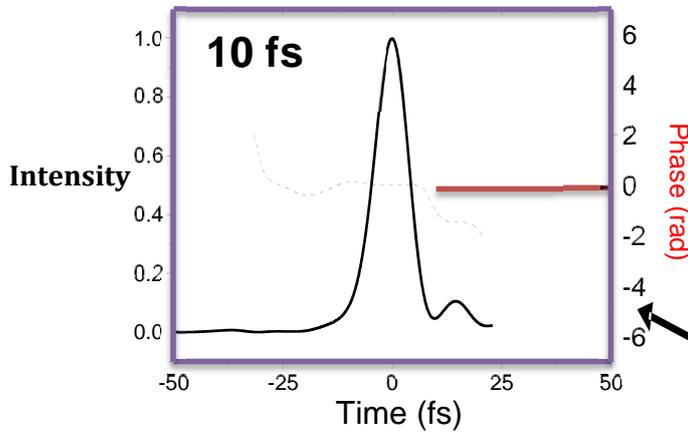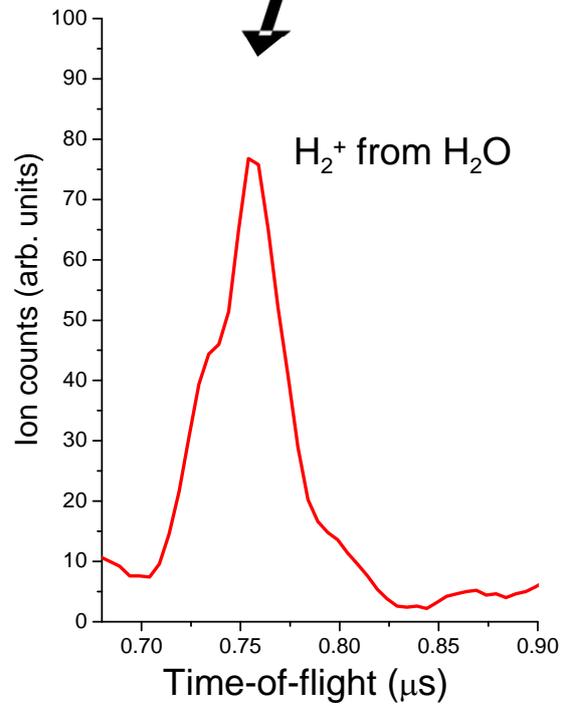

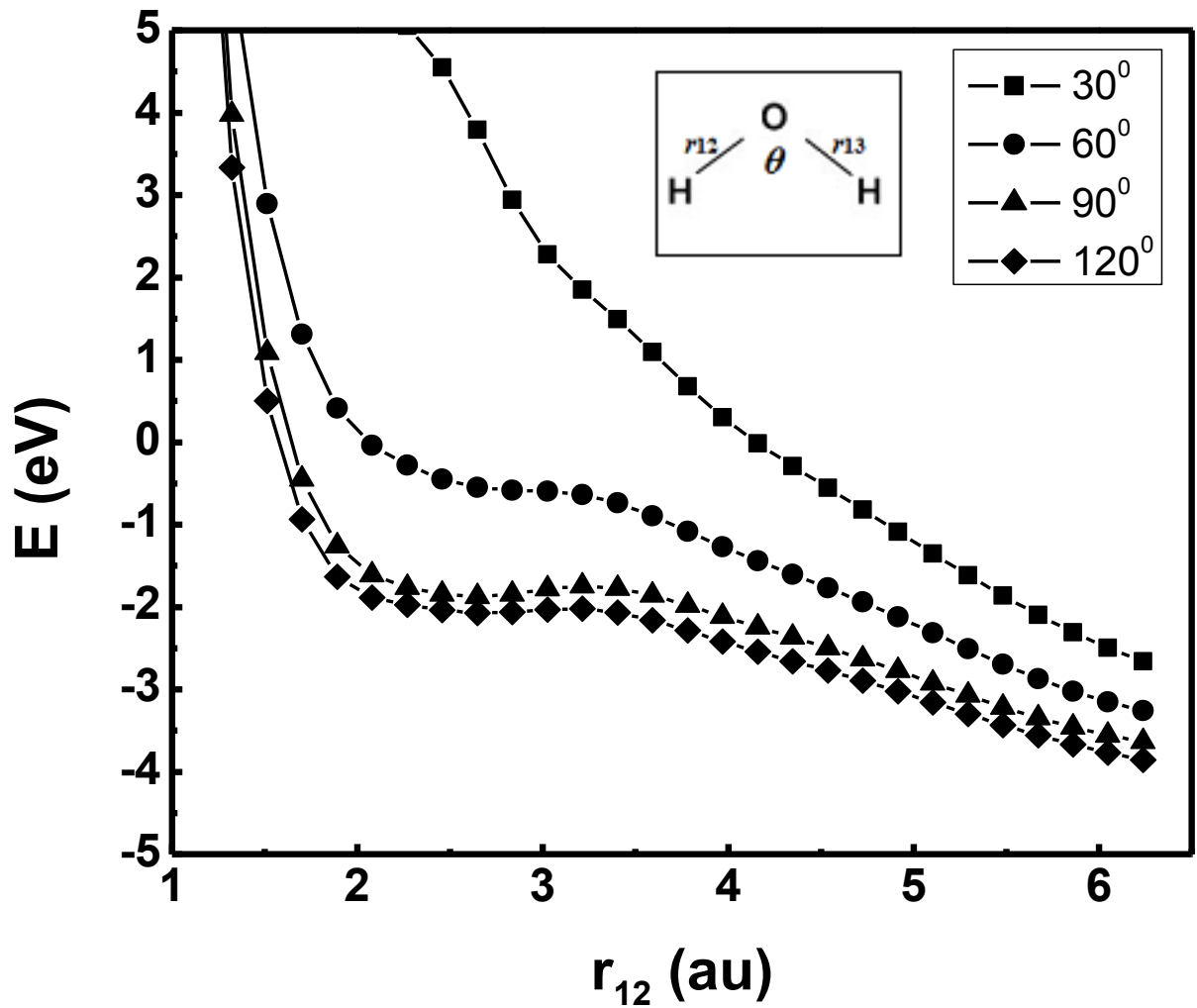

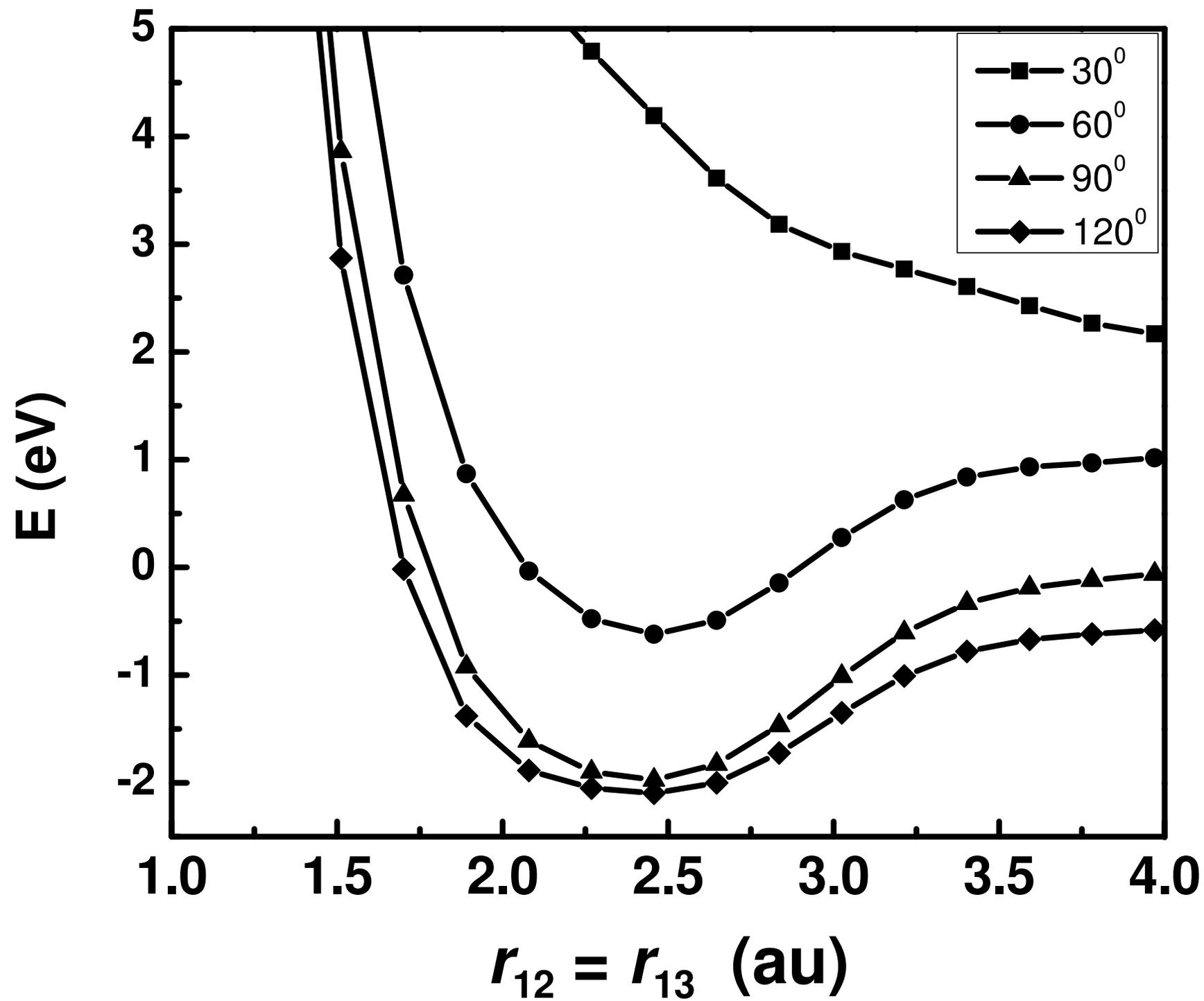

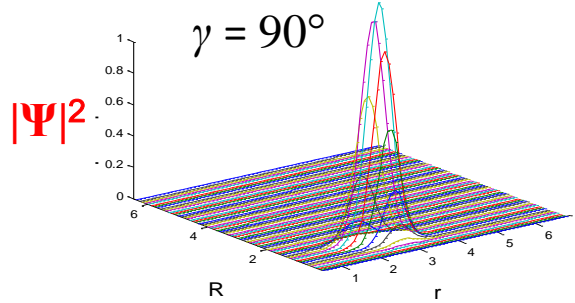 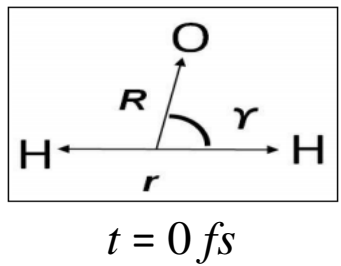 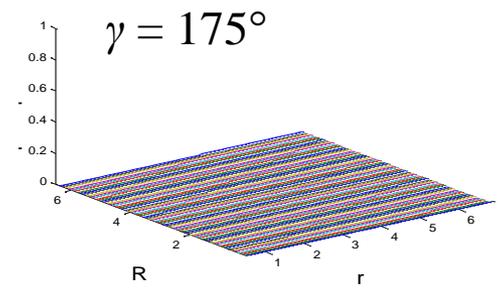

γ = 90°   t = 0 fs   γ = 175°

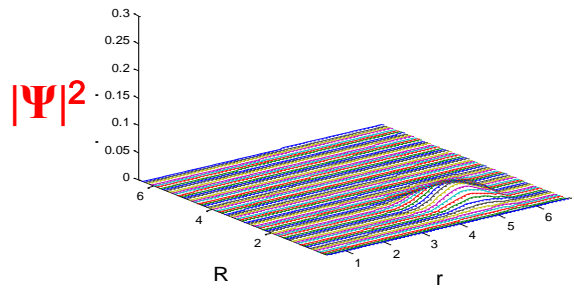 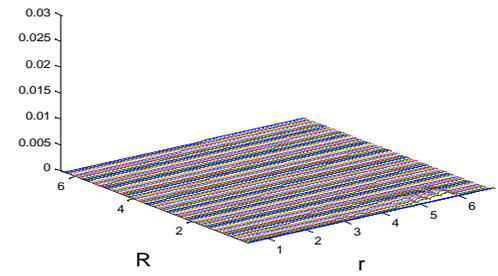

t = 5 fs

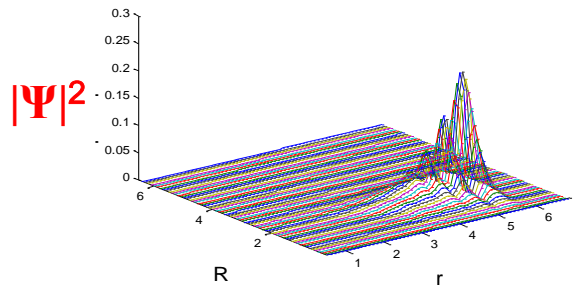 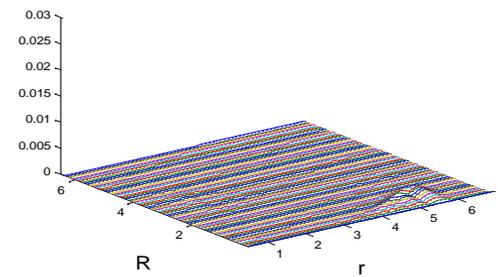

t = 10 fs

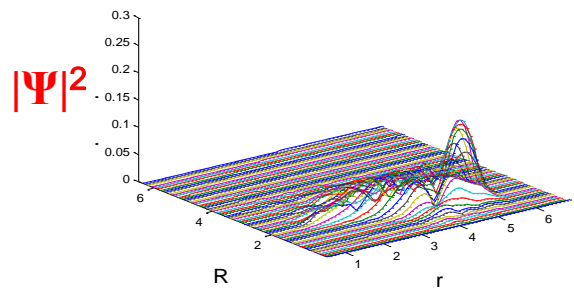 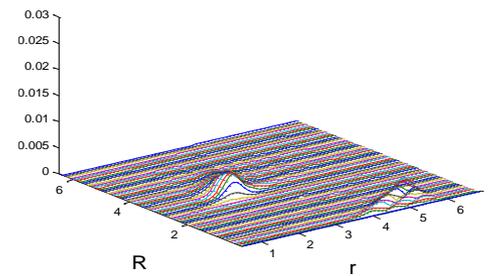

t = 15 fs

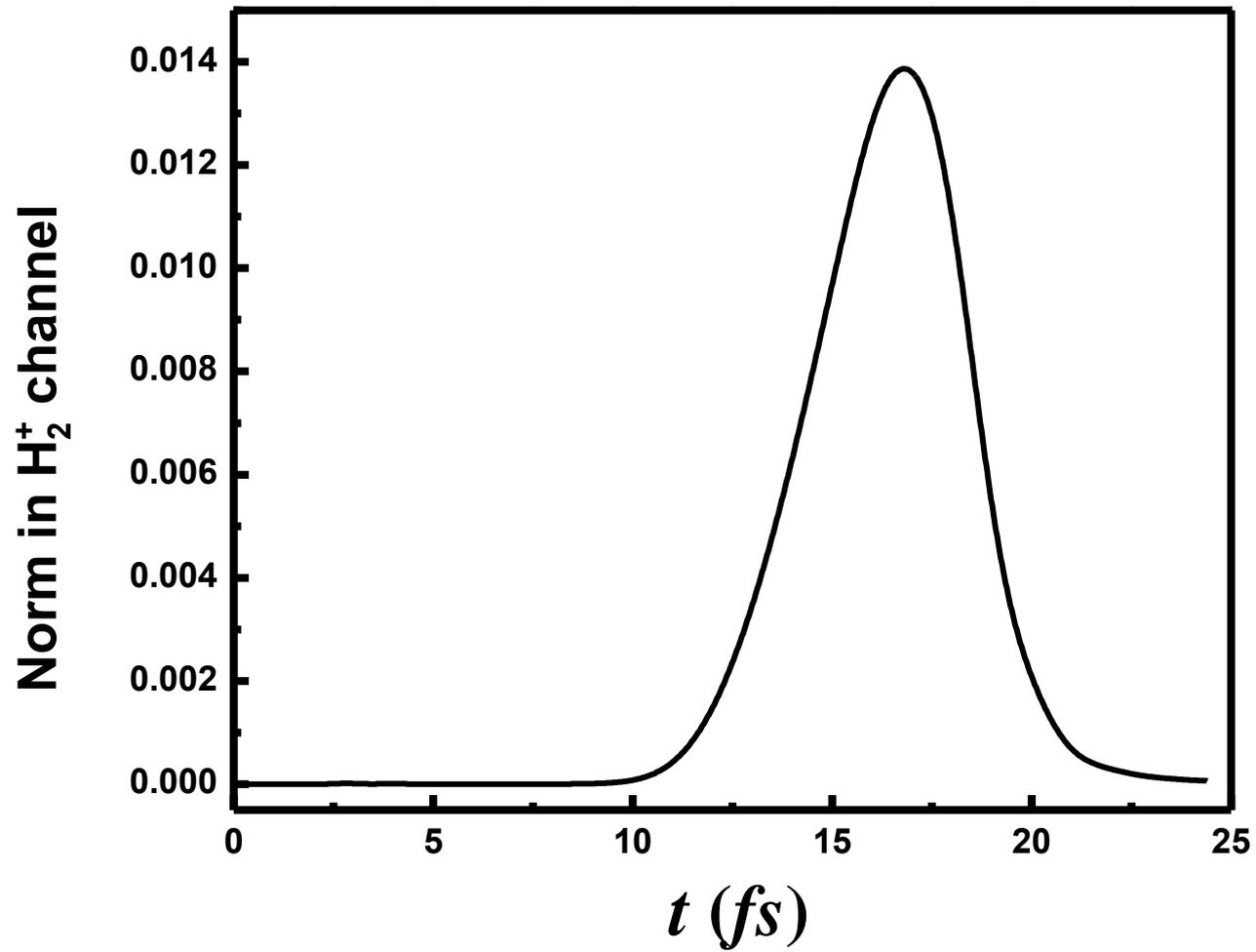

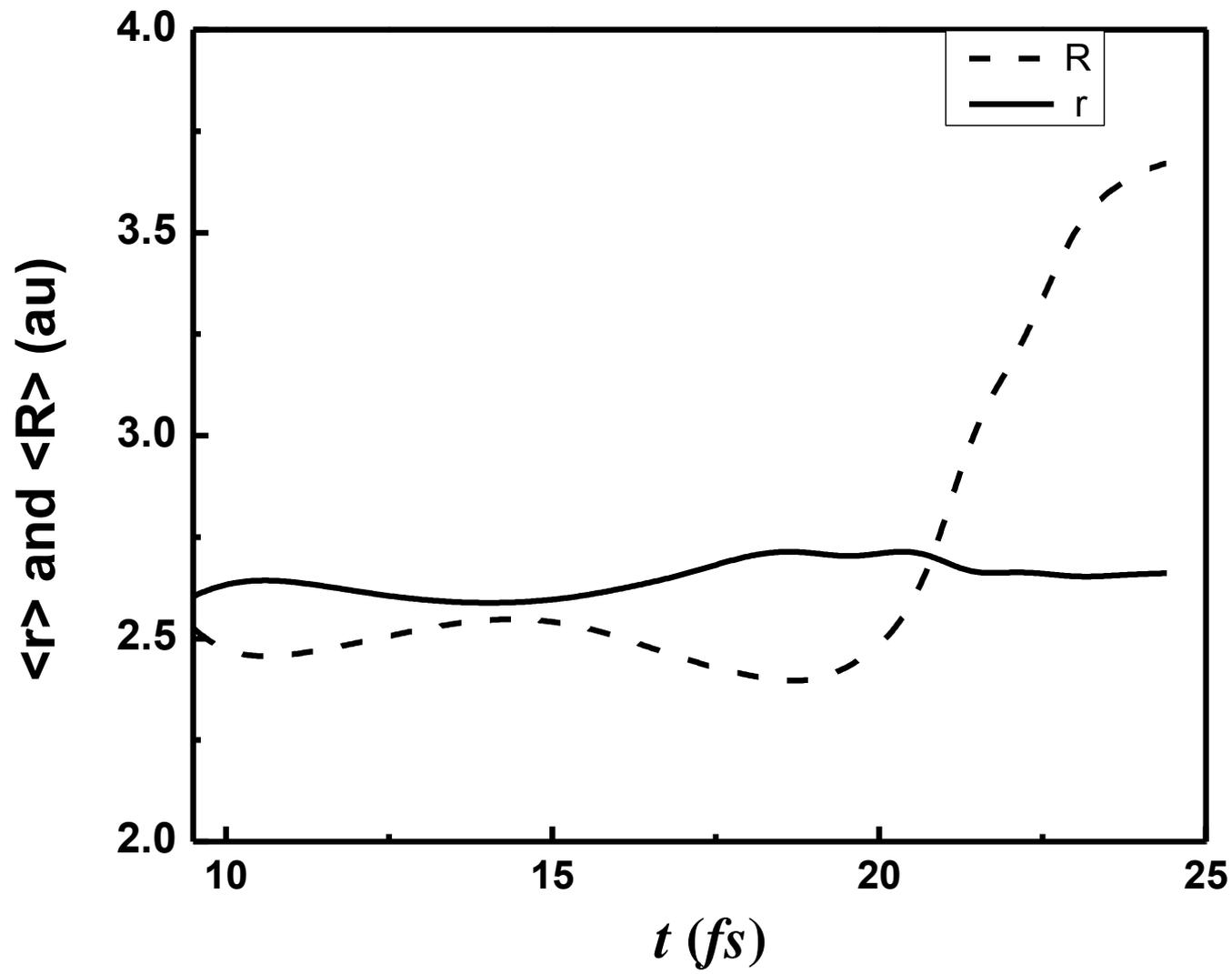